\documentclass[12pt,twocolumn]{aastex}

\usepackage{graphicx}
\usepackage{epsfig}
\usepackage{natbib}
\usepackage{hyperref}
\usepackage{bm}
\usepackage{amsmath}

\usepackage[normalem]{ulem}

\renewcommand{\vec}[1]{\bm{#1}}

\begin{document}

\title{A Statistical Inference Method for Interpreting the CLASP Observations}

\author{
J.~\v{S}t\v{e}p\'an\altaffilmark{1,2},
J.~Trujillo Bueno\altaffilmark{2,3,4},
L.~Belluzzi\altaffilmark{5,6},
A.~Asensio Ramos\altaffilmark{2},
R.~Manso Sainz\altaffilmark{7},
T.~del Pino Alem\'an\altaffilmark{2},
R.~Casini\altaffilmark{8},
R.~Kano\altaffilmark{9},
A.~Winebarger\altaffilmark{10},
F.~Auch\`ere\altaffilmark{11},
R.~Ishikawa\altaffilmark{9},
N.~Narukage\altaffilmark{9},
K.~Kobayashi\altaffilmark{10},
T.~Bando\altaffilmark{9},
Y.~Katsukawa\altaffilmark{9},
M.~Kubo\altaffilmark{9},
S.~Ishikawa\altaffilmark{12},
G.~Giono\altaffilmark{9},
H.~Hara\altaffilmark{9},
Y.~Suematsu\altaffilmark{9},
T.~Shimizu\altaffilmark{12},
T.~Sakao\altaffilmark{12},
S.~Tsuneta\altaffilmark{12},
K.~Ichimoto\altaffilmark{9,13},
J.~Cirtain\altaffilmark{14},
P.~Champey\altaffilmark{10},
B.~De~Pontieu\altaffilmark{16,17,18},
and
M.~Carlsson\altaffilmark{17,18}
}
\altaffiltext{1}{Astronomical Institute ASCR, Fri\v{c}ova 298, 251\,65 Ond\v{r}ejov, Czech Republic}
\altaffiltext{2}{Instituto de Astrof\'isica de Canarias, E-38205 La Laguna, Tenerife, Spain}
\altaffiltext{3}{Universidad de La Laguna, Departamento de Astrof\'isica, 38206 La Laguna, Tenerife, Spain}
\altaffiltext{4}{Consejo Superior de Investigaciones Cient\'ificas, Spain}
\altaffiltext{5}{Istituto Ricerche Solari Locarno, CH - 6605 Locarno Monti, Switzerland}
\altaffiltext{6}{Kiepenheuer-Institut f\"ur Sonnenphysik, D-79104 Freiburg, Germany}
\altaffiltext{7}{Max-Planck-Institut f\"ur Sonnensystemforschung, Justus-von-Liebig-Weg 3, 37077 G\"ottingen, Germany}
\altaffiltext{8}{High Altitude Observatory, National Center for Atmospheric Research, Post Office Box 3000, Boulder, CO 80307-3000, USA}
\altaffiltext{9}{National Astronomical Observatory of Japan, National Institutes of Natural Science, 2-21-1 Osawa, Mitaka, Tokyo 181-8588, Japan}
\altaffiltext{10}{NASA Marshall Space Flight Center, ZP 13, Huntsville, AL 35812, USA}
\altaffiltext{11}{Institut d'Astrophysique Spatiale, B\^atiment 121, Univ. Paris-Sud - CNRS, 91405 Orsay Cedex, France}
\altaffiltext{12}{Institute of Space and Astronautical Science, Japan Aerospace Exploration Agency, 3-1-1 Yoshinodai, Chuo, Sagamihara, Kanagawa 252-5210, Japan}
\altaffiltext{13}{Hida Observatory, Kyoto University, Takayama, Gifu 506-1314, Japan}
\altaffiltext{14}{University of Virginia, Department of Astronomy, 530 McCormick Road, Charlottesville, VA 22904, USA}
\altaffiltext{15}{University of Alabama in Huntsville, 301 Sparkman Drive, Huntsville, AL 35899, USA}
\altaffiltext{16}{Lockheed Martin Solar and Astrophysics Laboratory, Palo Alto, California, USA}
\altaffiltext{17}{Rosseland Centre for Solar Physics, University of Oslo, P.O. Box 1029 Blindern, NO-0315 Oslo, Norway}
\altaffiltext{18}{Institute of Theoretical Astrophysics, University of Oslo, P.O. Box 1029 Blindern, NO-0315 Oslo, Norway}

\shorttitle{Statistical Inference from Solar Spectropolarimetric Data}

\shortauthors{\v{S}t\v{e}p\'an, J. et al.}

\begin{abstract}
On 3rd September 2015, the Chromospheric Lyman-Alpha SpectroPolarimeter (CLASP) successfully measured the linear polarization produced by scattering processes in the hydrogen Lyman-$\alpha$ line of the solar disk radiation, revealing conspicuous spatial variations in the $Q/I$ and $U/I$ signals. Via the Hanle effect the line-center $Q/I$ and $U/I$ amplitudes encode information on the magnetic field of the chromosphere-corona transition region (TR), but they are also sensitive to the three-dimensional structure of this corrugated interface region.  With the help of a simple line formation model, here we propose a statistical inference method for interpreting the Lyman-$\alpha$ line-center polarization observed by CLASP.   

\end{abstract}

\date{Received XXXX; accepted XXXX}

\keywords{methods: statistical --- polarization --- radiative transfer --- Sun: chromosphere --- Sun: transition region}

\section{Introduction}
\label{sec:intro}

\begin{figure*}
\centering
\includegraphics[width=17cm]{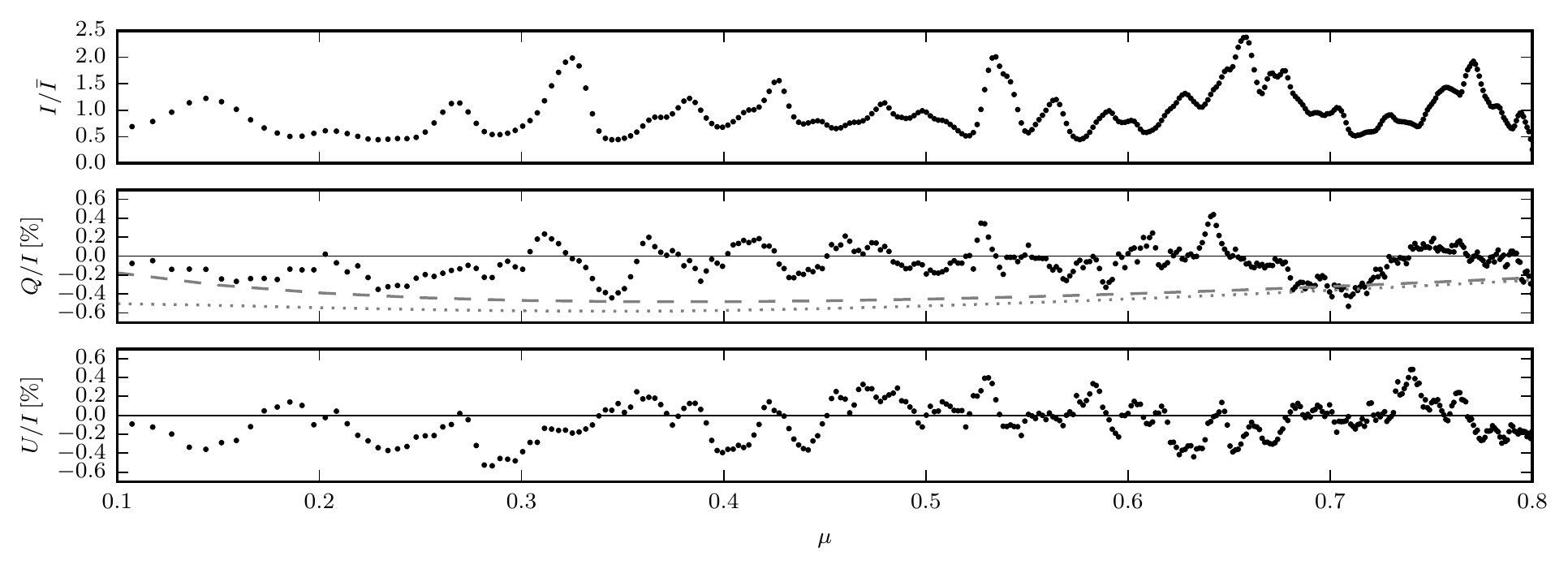}
\caption{
CLASP observations of the intensity (top panel), $Q/I$ (central panel), and $U/I$ (bottom panel) line-center signals of the hydrogen Ly-$\alpha$ line. 
Each point corresponds to a pixel along the slit of the spectrograph with its corresponding value for the cosine of the heliocentic angle, $\mu$. 
The reference direction for Stokes $Q$ positive is the parallel to the nearest limb.
The dashed curve in the central panel shows the center-to-limb variation of the $Q/I$ line-center signal calculated by \citet{jtb-stepan-casini11} in the semi-empirical model C of \citet{fontenla93}, while the dotted curve shows the center-to-limb variation of the ${\langle Q \rangle}/{\langle I \rangle}$ line-center signal calculated by \citet{stepanjtb15} in the 3D MHD model of \citet{carlsson16} (with the symbol ${\langle ... \rangle}$ meaning spatial average at each $\mu$ position).
We note that the standard deviation of the observed $Q/I$ and $U/I$ signals due to the presence of noise is $\sigma=0.05\,\%$, i.e., comparable to the vertical size of the data markers in the plot.
}
\label{fig:clasp}
\end{figure*}

Recently, the Chromospheric Lyman-Alpha SpectroPolarimeter (CLASP) sounding rocket experiment provided the first ever successful measurement of the linear polarization $Q/I$ and $U/I$ profiles produced by scattering processes in the hydrogen Ly-$\alpha$ line of the solar disk radiation \citep{kano17},  as well as in the Si {\sc iii} resonance line at 1206 \AA\ \citep{Ishikawa17}.  Such novel spectropolarimetric observations, with a spatial resolution of about 3~arcseconds, confirmed the following theoretical predictions for the Ly-$\alpha$ line \citep{jtb-stepan-casini11,belluzzi12,stepanjtb15}:
\begin{enumerate}
\item at the CLASP spatial resolution, the line-center $Q/I$ and $U/I$ signals, where the Hanle effect operates, are smaller than 1\% (see Fig.~\ref{fig:clasp});
\item the $Q/I$ and $U/I$ wing signals are larger than 1\%, with the $Q/I$ ones showing a clear center-to-limb variation (CLV) with negative amplitudes increasing towards the limb;    
\item both the line-core and wing signals show fluctuations along the spatial direction of the spectrograph's slit, which in the CLASP experiment was radially oriented from 20~arcseconds off-limb till 380~arcseconds on the solar disk.
\end{enumerate}

Although the CLASP observation of the hydrogen Ly-$\alpha$ line confirmed the above-mentioned theoretical predictions, it also revealed an interesting surprise, namely the lack of center-to-limb variation in the $Q/I$ line-center signal, in contrast with the predictions resulting from detailed radiative transfer calculations in one-dimensional (1D) semi-empirical and hydrodynamical models
\citep{jtb-stepan-casini11,belluzzi12} and in \citet{carlsson16} three-dimensional (3D) magneto-hydrodynamical (MHD) model of the solar atmosphere \citep{stepanjtb15}. 

The CLASP observations encode unique information on the 3D structure of the upper solar chromosphere. In particular, via the Hanle effect the line-center 
$Q/I$ and $U/I$ signals encode information on the magnetic field of the chromosphere-corona transition region (TR), but the same linear polarization signals are also sensitive to the geometry of this complex interface region. In order to constrain the mean field strength and the geometrical complexity of the chromosphere-corona TR from the CLASP observation, it is necessary to develop suitable inference methods. In this paper, we propose a statistical inference method for interpreting the observed Ly-$\alpha$ line-center polarization, and illustrate its applicability with the help of a simple radiative transfer model of the formation of the Ly-$\alpha$ line-core radiation in the corrugated surface that delineates the transition region. In our next paper (Trujillo Bueno et al. 2018; in preparation) the statistical inference method explained here will be applied to the CLASP line-center data in order to constrain the magnetic field and the geometrical complexity of the solar transition region via radiative transfer calculations in suitably parametrized three-dimensional models of the solar atmosphere.

\section{Ly-$\alpha$ Line Formation in a 3D Medium}
\label{sec:3d}

In semi-empirical models of the solar atmosphere, the core of the hydrogen Ly-$\alpha$ line forms in a thin (10--100\,km) layer at the base of the chromosphere-corona transition region. Below this layer, the anisotropy of the line-core radiation is practically zero \citep[see Fig.~1 of][]{jtb-stepan-casini11}; therefore, such deeper regions do not contribute to the scattering polarization at the line core. An analogous situation occurs in 3D models resulting from MHD simulations, such as that of \citet{carlsson16}. However, in this 3D model of the solar atmosphere the layer of line-core formation is no longer horizontally oriented, but a highly corrugated layer whose upper surface delineates the chromosphere-corona transition region \citep[see Fig.~3 of][]{stepanjtb15}. At each spatial point, the thickness of this corrugated layer is typicallly much smaller than the local radius of curvature of its surface; hence, one can look at the Ly-$\alpha$ line-core formation as of taking place locally in a thin planar slab located at the height of line-center optical depth unity, and having a well defined normal vector, $\vec n$, which points in the direction of the gradient of the kinetic temperature.\footnote{An alternative definition is possible, with $\vec n$ being the normal vector to the $\tau=1$ surface. Both definitions lead to very similar results, which provides an additional justification of the approximation.} The concept of local normal vector of the TR has already been introduced in Fig.~7 of \citet{stepanjtb15} in order to quantify the degree of corrugation of the TR.

The variation of the physical quantities along the local normal vector $\vec n$ is much more important than along the perpendicular direction. Therefore, it is a reasonable approximation to assume that the Ly-$\alpha$ line core originates in a plane-parallel atmosphere whose normal vector $\vec n$ is inclined with respect to the local vertical. This picture is useful because it allows us to use locally inclined 1D models to approximate the complex 3D structure of the chromosphere-corona transition region. We call this approximation the Corrugated Transition Region (CTR) model and we develop it quantitatively in Appendix~\ref{app:ctr}. 

As shown below, our simple CTR model is capable to relatively easily explain the surprising behavior of the Ly-$\alpha$ polarization observed by CLASP, namely that the $Q/I$ line-center signal does not show any clear CLV. In this paper, we use the CTR model to shed light on the sensitivity of the Ly-$\alpha$ line-center scattering polarization to the mean field strength and geometrical complexity of the TR and to illustrate the general statistical inference method we propose for interpreting the CLASP observations. We note, however, that the CTR model should be understood only as a rough approximation to the realistic 3D problem of the scattering polarization in the Ly-$\alpha$ line, because it fails at some fundamental levels if applied to the analysis of real spectropolarimetric observations (see Sect.~\ref{ssec:ctrlim}). In other words, in this paper the CTR model is used only for illustrative purposes and for facilitating the explanation of our statistical inference method, but not for interpreting any real observation of the Ly-$\alpha$ scattering polarization. As we shall show in our next paper, our interpretation of the CLASP data is instead based on a statistical model of the solar TR that uses many 3D models of the solar atmosphere, which we have constructed starting from a state-of-the-art model resulting from a MHD simulation.

Finally, it is necessary to emphasize that while calculations of the line intensity using the so-called 1.5D approximation, by means of which the radiation transfer in each column of the model atmosphere is treated as if it were a 1D plane-parallel atmosphere, can sometimes be a sufficiently good approximation, if scattering polarization is considered, the 1.5D approximation often fails \citep[e.g.,][]{stepan-jtb16}.

\subsection{Center-to-limb Variation and the Symmetry Breaking Problem}

\begin{figure*}
\centering
\includegraphics[width=17cm]{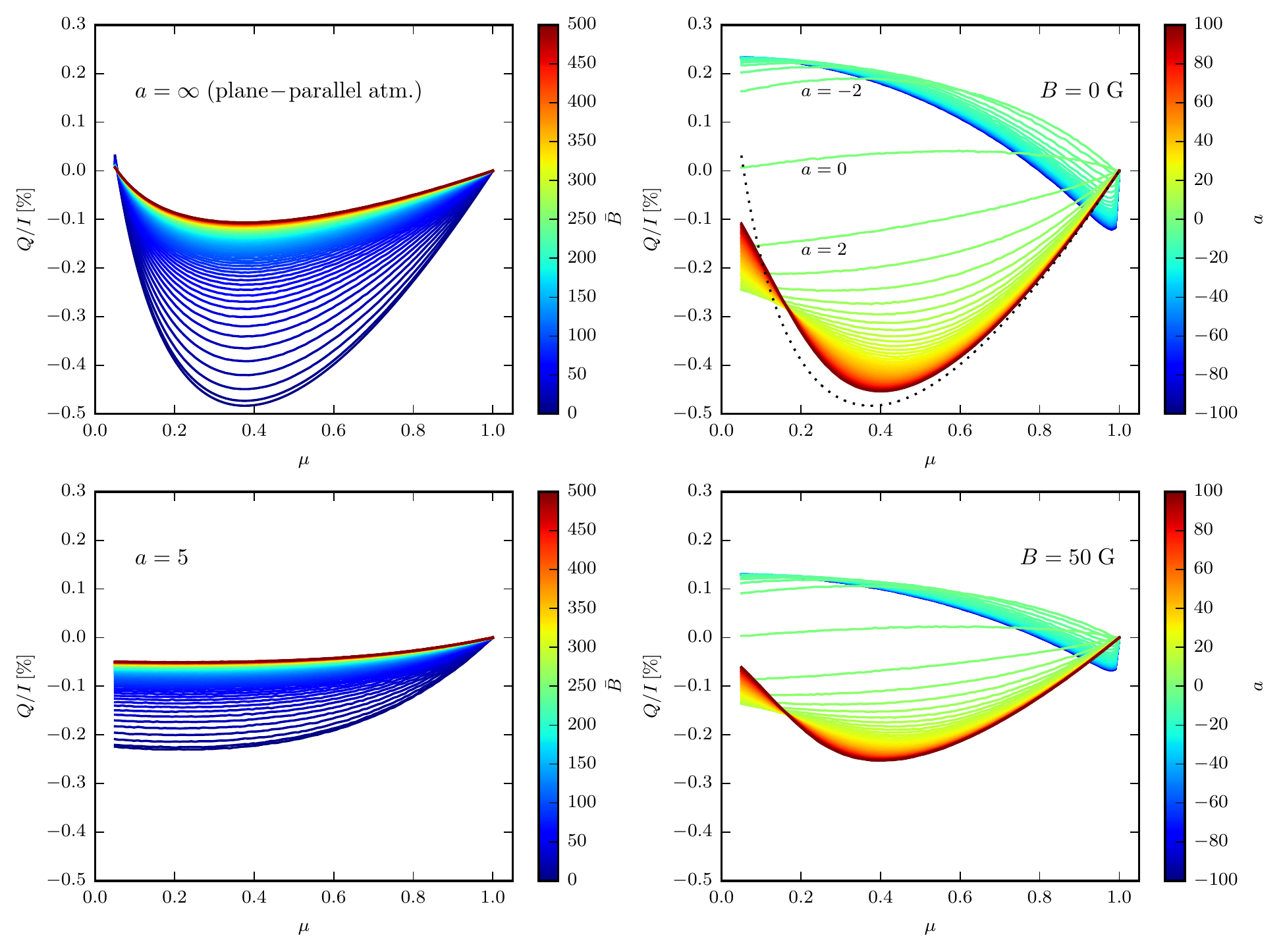}
\caption{CLV of the Ly-$\alpha$ $Q/I$ line-center signals calculated in several CTR models.
Top left panel: the case of the plane-parallel FAL-C model ($a=\infty$) with increasing mean field strengths $\bar B$. Top right panel: unmagnetized CTR models ($\bar B=0\,{\rm G}$) with various corrugation parameters $a$. The black dotted line shows the CLV of the original FAL-C model, $(\infty,0)$. Bottom left panel: CTR models with $a=5$ and increasing $\bar B$ values. Bottom right panel: CTR magnetized model ($\bar B=50$\,G) with various corrugation parameters $a$. The step between the curves corresponding to different $\bar B$ values is 5\,G, while the step between the curves with different corrugation parameters $a$ is 2.
}
\label{fig:clvs}
\end{figure*}

\begin{figure*}
\centering
\includegraphics[width=17cm]{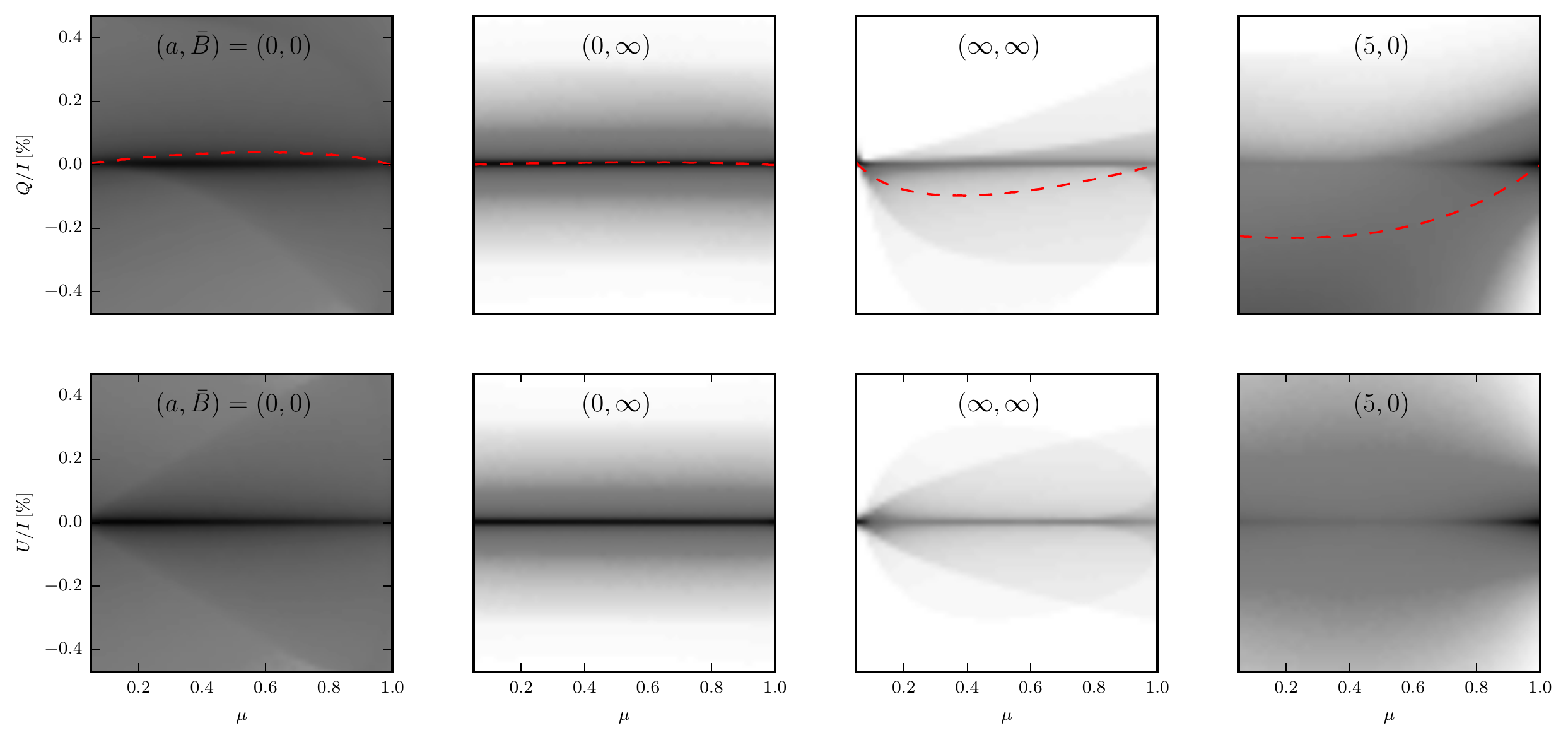}
\caption{
Probability density functions $p^\mu(q|\vec\theta)$ (top panels) and $p^\mu(u|\vec\theta)$ (bottom panels) for various CTR model parameters $\vec\theta=\{a,\bar B\}$ (indicated at the top of each panel). 
For the sake of simplicity, we assume that the joint PDF can be factorized as $p^{\mu}(\vec S|\vec\theta)=p^{\mu}(q|\vec\theta)p^{\mu}(u|\vec\theta)$.
For each value of $\mu$ (horizontal axis), the darker shades of grey indicate larger PDF values.
The dashed red curves in the top panels show the CLV of the $q$ signals after averaging the $q$ values at each $\mu$ (cf., Fig.~\ref{fig:clvs}). The PDFs are normalized to unity at each $\mu$ value. See the text for more details. The color table differs from panel to panel in order to improve the contrast of the individual plots. The normalization is such that $\int p^\mu(q|\vec\theta)\;dq=\int p^\mu(u|\vec\theta)\;du=1$ for every $\mu$.
}
\label{fig:ctrpdfdist}
\end{figure*}

The linear polarization of the emergent spectral line radiation is determined by the atomic polarization induced in the line's levels by the pumping radiation field, and by the orientation of the line of sight. The atomic polarization of a level of angular momentum $J$, which can be suitably quantified by the $\rho^K_Q$ multipolar components of the atomic density matrix, are closely connected to the radiation field anisotropy in the line-formation region. The symmetry properties of the radiation field can be conveniently described in terms of the irreducible radiation field tensor $J^K_Q$ \citep{LL04}. These quantities are determined by the thermodynamical and dynamical structure of the atmosphere. In the idealized case of a static, unmagnetized 1D plane-parallel model atmosphere, the radiation field is fully determined by the vertical stratification of the atmosphere and, due to the cylindrical symmetry of the problem, the only non-zero components of the radiation field tensor are the familiar mean intensity, $J^0_0$, and the anisotropy component $J^2_0$. In such a model atmosphere, the linear polarization of the emergent spectral line radiation can be either parallel or perpendicular to the projection of the local vertical (symmetry axis of the problem) onto the field of view (FOV). For more information on the anisotropy and polarization of the hydrogen Ly-$\alpha$ line in plane-parallel models of the solar atmosphere, see Sect.~3 of \citet{jtb-stepan-casini11}.

In the CTR model the local normal vector $\vec n$ of the TR can be inclined with respect to the local vertical, and the reference direction of linear polarization changes accordingly, being either parallel or perpendicular to the projection of $\vec n$ onto the FOV (because the local symmetry axis of the radiation field is now $\vec n$ instead of $Z$). The modification of the emergent polarization by deviations of the model's physical conditions from the 1D plane-parallel limit is usually referred to as the symmetry breaking effect \citep[e.g.,][]{mansojtb11} and it has been demonstrated, using self-consistent 3D simulations, that it affects the polarization signals of many chromospheric lines including Ly-$\alpha$ \citep[e.g.,][]{stepan15,stepanjtb15,stepan-jtb16}.

The symmetry breaking effects pose a great challenge for the interpretation of the scattering polarization spectra because the magnitude and orientation of the spectral line polarization can be misinterpreted as being due to the action of a magnetic field via the Hanle effect. \citet{carlin12} emphasized that the gradients of the vertical component of the macroscopic velocity can change the polarization amplitudes, while \citet{stepan-jtb16} pointed out that the horizontal components of the macroscopic velocity can have an important impact on the scattering polarization signals via their breaking of the axial symmetry of the pumping radiation field. Fortunately, for the particular case of the Ly-$\alpha$ line, the impact of the dynamical state of the model atmosphere is not very important, due to the large Doppler width of Ly-$\alpha$ and to its line-core formation taking place within a geometrically thin layer.

The CTR model allows us to construct random realizations of the TR at any given point on the solar disk, characterized by the cosine of the heliocentric angle, $\mu=\cos\theta$. In each such realization, the TR normal vector has a random inclination (sampled from the distribution given by Eq.~\ref{eq:nvdist}) and the magnetic field vector has a random orientation and magnitude sampled from the distribution of Eq.~(\ref{eq:bdist}). If we average the Stokes $I$, $Q$, and $U$ signals emerging from many such random realizations, we can construct the center-to-limb variation curves which provide some information about the average properties of the atmosphere (hereafter, average CLV curves).

The impact of different levels of magnetization, quantified by the $\bar B$ parameter of Eq. (A4), and corrugation, quantified by the $a$ parameter of Eq. (A3), can be seen in the CLV curves of $Q/I$ shown in Fig.~\ref{fig:clvs}. The simplest case of a non-magnetized 1D plane-parallel atmosphere corresponds to the CLV curve with the largest amplitude, shown in the top left panel of the figure. As the magnetic field increases, the polarization amplitudes are reduced by the Hanle effect. In the limit of very strong fields \citep[the so-called Hanle effect saturation regime;][]{LL04} the polarization amplitudes are about a factor $1/5$ lower than in the non-magnetized atmosphere.

The case of a slightly corrugated atmosphere ($a=5$) is shown in the bottom left panel. As in the 1D plane-parallel model, we see the depolarizing effect of a magnetic field of increasing strength, up to the Hanle saturation limit.

The top right panel of Fig.~\ref{fig:clvs} demonstrates the effect of different corrugations of a non-magnetized atmosphere. In the limit of a very large positive $a$ value, we obtain the plane-parallel solution. As the corrugation parameter $a$ decreases, the $Q/I$ amplitudes also decrease. At the same time, the near-limb $Q/I$ values tend to increase for an interval of $a$ values. This is due to the fact that the sharp $Q/I\approx 0$ value around $\mu=0.1$ in the plane-parallel case gets mixed with larger $Q/I$ signals from models having slightly different orientations of the local TR normal vector. A similar behavior can be seen in more realistic investigations based on 3D models \citep[see Fig.~12 of][]{stepanjtb15}. For $a=0$, the orientation of the TR normal vectors is isotropic in the upper hemisphere. This case corresponds to the maximum level of corrugation of the atmosphere and the corresponding CLV practically vanishes.\footnote{The small CLV residual is due to the fact that the inclination of the normal vectors in the CTR model is restricted to the $[0^\circ,90^\circ]$ interval.} For $a<0$ we see that the CLV curve is mostly positive, although it shows a sign reversal around $\mu{=}0.8$. An increasingly negative corrugation parameter makes the local TR normal vectors increasingly predominantly horizontal. This situation corresponds to the case in which the plasma structures in the model atmosphere are mostly vertically oriented, such as in the case of solar spicules. In the limit $a\to-\infty$, the normal vectors become horizontal. It is in this extreme case that we obtain the largest negative $Q/I$ values near the disk center ($\mu\approx1$). We note, however, that in the limit $\mu\to 1$ the $Q/I$ value is zero.

In the bottom right panel of Fig.~\ref{fig:clvs}, we show a case analogous to that of the upper right panel, but here with an average strength $\bar B=50$\,G, which is close to the critical Hanle field of Ly-$\alpha$. The impact of the magnetic field is simply a depolarization of the signals.

\subsection{Distribution of the Polarization Signals and Statistical Models of the Atmosphere}
\label{ssec:distrps}

Hereafter, we will often use $q=Q/I$, $u=U/I$ and $\vec S=[q,u]^{\rm T}$ in order to simplify the notation. 

As seen in Fig.~\ref{fig:clasp}, the $q$ and $u$ line-center signals observed by CLASP are fluctuating around zero with typical amplitudes around 0.2\,\%. The Doppler core of the hydrogen Ly-$\alpha$ line forms in the upper chromosphere, and it is important to note that the Hanle effect operates at the line center, where the CRD approximation provides a suitable approximation \citep[see][]{belluzzi12}. The linear polarization of this resonance line is due to scattering processes in a complex 3D medium and to the action of the Hanle effect. The Zeeman effect can be neglected due to the large Doppler width of the line and the weakness of the magnetic field in the upper chromosphere of the quiet Sun. Since the hydrogen Ly-$\alpha$ line is very broad, the scattering polarization produced by anisotropic radiation pumping in this 
line is not very sensitive to the Doppler shifts produced by the macroscopic motions of the plasma. This contrasts with the case of narrower chromospheric lines whose linear polarization signals can be significantly affected by the presence of macroscopic velocity field gradients \citep[see, e.g.,][]{stepan-jtb16}. An {\em estimate} of the average CLV of the $q$ signals observed by CLASP by, for instance, a parabolic fit of the data, gives an almost zero CLV, which is in sharp contrast with the results obtained using the above-mentioned semi-empirical and 3D~MHD models (see the dashed and dotted lines of Fig. \ref{fig:clasp}).

In general, the statistical models chosen for interpreting the CLASP observations should take into account the fact that there may exist correlations between the intensity and polarization signals. This has been predicted by \citet{stepanjtb15} and later confirmed by CLASP \citep[see][]{kano17}, and in Trujillo Bueno et al. (2018) we will take into account that plasma structures, such as internetwork and network regions, differ not only in intensity but also in their polarization signals. However, for the sake of simplicity, in this paper, we ignore such correlations. This is natural in the case of our simple CTR model, which is based on a single semi-empirical model atmosphere with its corresponding line-center intensity at any given $\mu$ value. 

While a fitting of the data with a smooth CLV curve (e.g., with the parabolic fit previously mentioned) can be visually compared with the synthetic CLVs, such approach is not sufficient for a quantitative analysis. The main reason is that by fitting just the CLV we ignore a significant amount of information contained in the $q$ and $u$ data. If the average CLV of $q$ is zero, then all the models with zero average CLV would be considered to be equally suitable, regardless of the amplitude of local polarization fluctuations.

Let us assume that the quiet Sun atmosphere can be regarded as a stochastic system with observables having a definite probability distribution function at any given time, point on the FOV, and line of sight. The observable quantities of interest are the line-center Stokes parameters of the spectral line. In general, the statistical properties of the observables may depend on the phase of the solar cycle and on the solar latitude of the quiet Sun. It is important to clarify that we do not assume the so-called microturbulent limit in this paper. The magnetic field vector and the orientation of the transition region normal vector have definite values at any point of the FOV. However, these quantities can be considered to have random values because the plasma structures of solar disk are randomly oriented over the field of view and as time goes by.

Consider the idealized case of a spatially and temporally resolved observation. The observed fractional linear polarization signals at a given $\mu$ value result from a particular realization of the atmospheric corrugation and magnetic field at a given point and time. Let us assume that we have a statistical model of the quiet solar atmosphere depending on a finite set of parameters $\vec\theta$ common to all the points in the FOV.\footnote{In the hierarchical Bayesian formulation of the inference problem described below, we can consider these parameters as the hyperparameters of the model. Following the Bayesian terminology, the term hyperparameter stands for the parameter of the prior distribution \citep{GelmanHill07}. } These parameters define the global properties of the model atmosphere.

There is an associated probability density function $p^\mu(\vec S|\vec\theta)$ that quantifies the probability of $\vec S$ conditional on the truth of 
$\vec\theta$, which can be obtained from the histograms of the Stokes signals produced by the model. 
In the simple case of the CTR model, the set of model parameters is $\vec\theta=\{a,\bar B\}$ and the corresponding signals are obtained simply by generating random orientations and magnetic field strengths of the TR from the model's probability density functions controled by $\vec\theta$ (see Appendix~\ref{app:ctr}).
In general, the model can result from a 3D numerical simulation of the solar atmosphere, as we will show in our next paper.

In Fig.~\ref{fig:ctrpdfdist}, we show some examples of such PDFs for different model parameters. The general properties of these PDFs can be summarized as follows: 

(1) The $u$ PDFs are symmetric with respect to the zero value. This follows from the definition of the Stokes parameters and the symmetry of the problem ($u$ and $-u$ must be equally likely). 

(2) The $q$ PDFs are generally asymmetric, except at $\mu=1$ where they are equal to the $u$ PDFs.

The left panels of Fig.~\ref{fig:ctrpdfdist} correspond to the most corrugated ($a=0$) unmagnetized ($\bar B=0$) model. The CLV indicated by the red dashed line is very small but the spread of possible values of both $q$ and $u$ around zero is very significant for all LOS $\mu$ values. This is due to the fact that the relative orientation of the TR normal vectors with respect to the LOS is only weakly dependent on the LOS. As mentioned in the previous section, there is a remaining asymmetry in the problem due to the restriction of the TR normal vectors to the upper hemisphere, which causes a non-zero CLV and generally a dependence of $p^{\mu}$ on $\mu$.

The panels in the second column correspond to the same level of corrugation ($a=0$) but now with magnetic field in the saturation regime of the Hanle effect. The depolarizing effect of the magnetic field leads to a significant reduction of the probability of having large polarization signals. The $q$ and $u$ signals generated from this model would therefore have much smaller amplitudes along the spatial direction of the spectrograph slit, assuming it is radially oriented and extending from $\mu=0.1$ to $1$.

The third column of Fig.~\ref{fig:ctrpdfdist} shows the PDFs of a strongly magnetized FAL-C plane-parallel model atmosphere.
In this model, the symmetry breaking effects are only due to the action of the magnetic field. In contrast to the unmagnetized FAL-C model, in which the PDF would coincide with the red dashed curve of the upper panel of the figure, the magnetic field can produce both negative and positive $q$ and $u$ signals. While positive and negative $q$ values are equally likely near the disk center, there is a significant predominance of the negative signals for $\mu<1$. In contrast to these models, the $q$ signals observed by CLASP along the spectrograph slit show a symmetric distribution around zero (see the middle panel of Fig.~\ref{fig:clasp}).

The last column of Fig.~\ref{fig:ctrpdfdist} contains another example of the PDFs in a slightly corrugated unmagnetized model with predominantly vertical normal vectors ($a=5$). While there is a clear average CLV in $q$, the spread of values around it and the distribution of $u$ signals around zero indicates that the symmetry breaking mechanism plays a very significant role. This model resembles the case of the 3D\,MHD model of \citet{carlsson16} (see Appendix~\ref{app:ctr}).

Different combinations of magnetic fields and corrugation parameters produce a rich variety of probability density functions that contain much more information than the CLV curves. Instead of a single value of $q$ at a given $\mu$, one has PDFs for $q$ and $u$ that can be quantitatively compared with the distribution of observed data. This is important especially for observations like CLASP, where we have only a small amount of data. In the following section, we formulate the statistical inference method in a more rigorous way.

\section{The Statistical Inference Method}
\label{sec:stat}

\subsection{Bayesian Analysis}

We denote the observed line-center $Q/I$ and $U/I$ signals at the spatial position $i$ along the CLASP spectrograph slit 
by $\vec S^{\rm obs}(\mu_i)=[q^{\rm obs}(\mu_i),u^{\rm obs}(\mu_i)]^{\rm T}$, where $\mu_i$ is the cosine of the heliocentric angle.
Likewise, $\vec S(\mu_i;\vec\theta)=[q(\mu_i;\vec\theta),u(\mu_i;\vec\theta)]^{\rm T}$ represent the random $Q/I$ and $U/I$ line-center signals sampled from our model atmosphere characterized by the hyperparameters $\vec\theta$. For example, in the CTR model $\vec\theta=\{a,\bar B\}$.
Our generative model of the CLASP observations can be defined as follows: 

\begin{equation}
\vec S^{\rm obs}(\mu_i)=\vec S(\mu_i;\vec\theta)+\vec\epsilon\,,
\end{equation}
where $\vec\epsilon$ represents uncorrelated Gaussian noise with variance $\sigma^2$.

Using the notation of \citet{gregory05}, Bayes theorem reads
\begin{equation}
p(H_j|D,I)=\frac{p(H_j|I)p(D|H_j,I)}{p(D|I)}\,,
\label{eq:bayes}
\end{equation}
where for the case of the CLASP experiment we have
\begin{itemize}
\item $D\equiv \vec S^{\rm obs}$: proposition representing the data
\item $H_j\equiv \vec S,\vec\theta$: proposition asserting the truth of the model hypothesis 
\item $I$: proposition representing our prior information (e.g., the $\mu_i$ value, the PSF of the instrument, etc.)
\end{itemize}
The quantity $p(H_j|D,I)$ is the desired posterior probability density function of the hypothesis given the observed data, $p(H_j|I)$ is the prior probability distribution of the model parameters, and $p(D|H_j,I)$ is the likelihood, i.e., a measure of how well the given model can fit the observed data.

In the following, we drop $I$ from the equations for notational simplicity because it represents an obvious information we have about our measurements. Finally, we note that the denominator of Eq.~(\ref{eq:bayes}), the so-called evidence, satisfies the normalization condition
\begin{equation}
p(D|I)=\sum_j p(H_j|I)p(D|H_j,I)
\end{equation}
which ensures that $\sum_j p(H_j|D,I)=1$. Since this normalization constant is not relevant for the final expressions given below, hereafter we take it equal to unity.

\begin{figure*}
\centering
\includegraphics[width=17cm]{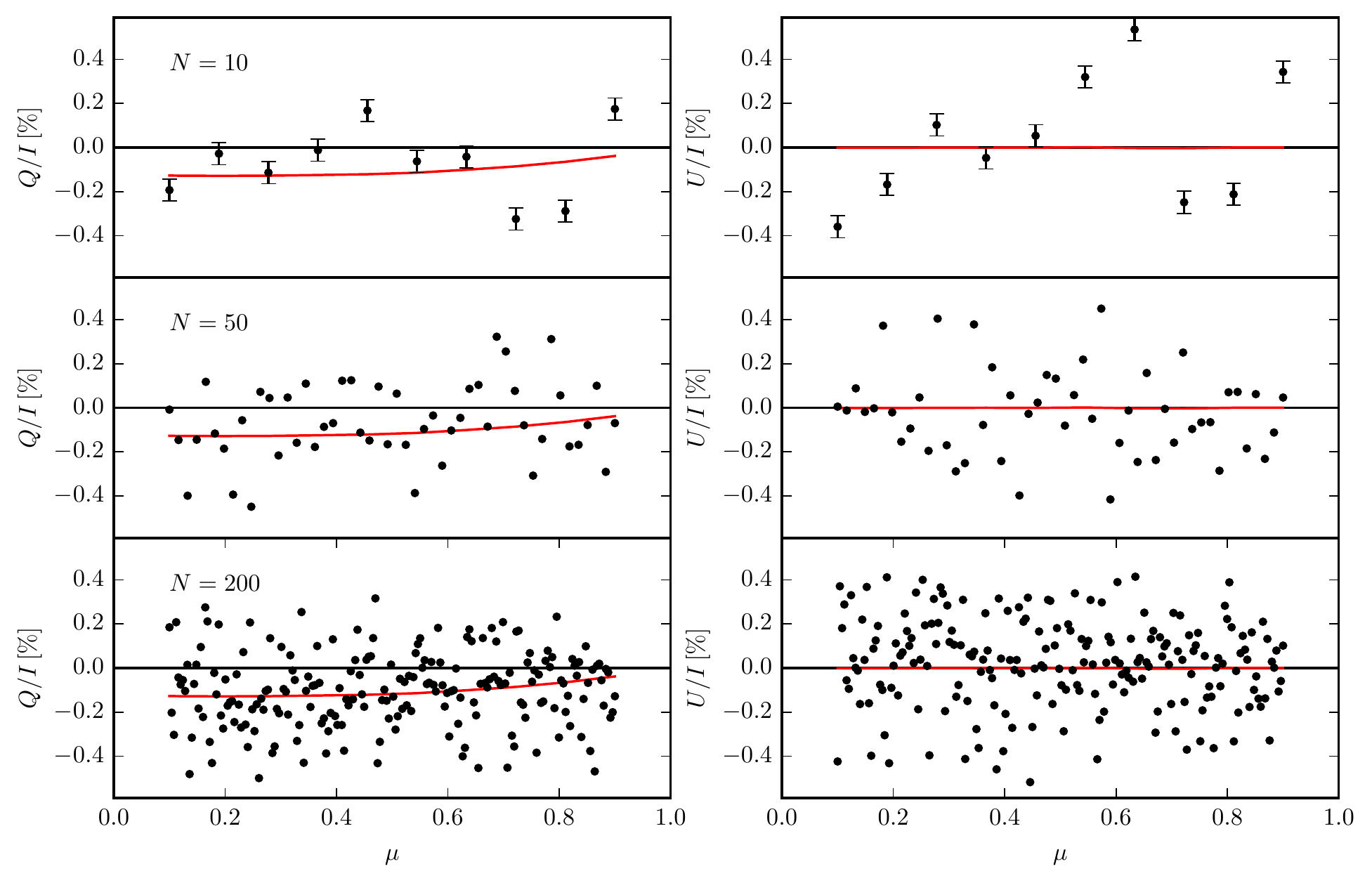}
\caption{
Multiple random realizations of the line-center $Q/I$ (left panels) and $U/I$ (right panels) signals of the Ly-$\alpha$ line 
sampled from the CTR model with parameters $\vec\theta_0=\{a_0,\bar B_0\}=\{5,50\,{\rm G}\}$. For an increasing number of data points, $N$, (i.e., the number of pixels along the imaginary spectrograph slit extending from $\mu=0.1$ to 0.9), the fractional polarization signals are distributed around the average CLV curves (shown with the red  curves) according to the probability density function $P^{\mu_i}_0(q_i,u_i|\vec\theta_0)$, where $i=1,\dots,N$.
We have added gaussian noise with standard deviation $\sigma=0.05\,\%$ to the data (the 1$\sigma$ bars are only shown in the top panels).
}
\label{fig:psalc}
\end{figure*}

In our notation, the Bayes theorem applied to each spatial pixel (characterized by $\mu_i$) of the CLASP spectrograph slit establishes 
that\footnote{For notational simplicity, we drop the explicit dependence of $\vec S$ and $\vec S^{\rm obs}$ on $\mu_i$.}

\begin{equation}
p^{\mu_i}(\vec S,\vec\theta|\vec S^{\rm obs}) \,=\, p^{\mu_i}(\vec S,\vec\theta) p^{\mu_i}(\vec S^{\rm obs}|\vec S,\vec\theta)\,.
\end{equation}
We now marginalize this posterior probability density function by integrating over $\vec S$, which allows us to obtain the following expression for the posterior of the hyperparameters:

\begin{eqnarray}
p^{\mu_i}(\vec\theta|\vec S^{\rm obs})&=&
\int d\vec S\;p^{\mu_i}(\vec S,\vec\theta|\vec S^{\rm obs})\nonumber\\
&=&\int d\vec S\; p^{\mu_i}(\vec S,\vec\theta) p^{\mu_i}(\vec S^{\rm obs}|\vec S,\vec\theta)\,.
\end{eqnarray}
The prior distribution $p^{\mu_i}(\vec S,\vec\theta)$ can be factorized as \citep[cf.][]{gregory05}
\begin{equation}
p^{\mu_i}(\vec S,\vec\theta)=p^{\mu_i}(\vec S|\vec\theta)p(\vec\theta)\,,
\end{equation}
where $p(\vec\theta)$ is the prior of the hyperparameters.
Under the assumption of uncorrelated Gaussian noise, the likelihood factor $p^{\mu_i}(\vec S^{\rm obs}|\vec S,\vec\theta)$ reads
\begin{eqnarray}
p^{\mu_i}(\vec S^{\rm obs}|\vec S,\vec\theta)=
\frac{1}{(\sqrt{2\pi}\sigma)^2}
\exp\left\{ -\frac{\| \vec S-\vec S^{\rm obs} \|^2}{2\sigma^2} \right\}
\end{eqnarray}
where we have used the compact notation
$\| \vec S-\vec S^{\rm obs} \|^2=(q-q^{\rm obs})^2+(u-u^{\rm obs})^2$. Finally, the marginal posterior of the hyperparameters 
for a single pixel reads
\begin{eqnarray}
p^{\mu_i}(\vec\theta|\vec S^{\rm obs})\,=\,\nonumber\\
\left[ \int d\vec S\;\frac{p^{\mu_i}(\vec S|\vec\theta)}{(\sqrt{2\pi}\sigma)^2}
\exp\left\{ -\frac{\| \vec S-\vec S^{\rm obs} \|^2}{2\sigma^2} \right\} \right]p(\vec\theta)\;.
\label{eq:poster}
\end{eqnarray}
By using the notation for the likelihood
\begin{equation}
\mathcal{L}_i(\vec\theta|\vec S^{\rm obs})\equiv\int d\vec S\;\frac{p^{\mu_i}(\vec S|\vec\theta)}{(\sqrt{2\pi}\sigma)^2}
\exp\left\{ -\frac{\| \vec S-\vec S^{\rm obs} \|^2}{2\sigma^2} \right\}\,,
\label{eq:llii}
\end{equation}
we can finally write the posterior of the hyperparameters based on the data from all the CLASP pixels:
\begin{equation}
p(\vec\theta|\vec S^{\rm obs})=\prod_i \mathcal{L}_i(\vec\theta|\vec S^{\rm obs}) p(\vec\theta)=\mathcal{L}(\vec\theta|\vec S^{\rm obs})p(\vec\theta)\,.
\label{eq:finpost}
\end{equation}

It is important to emphasize that in our forthcoming paper regarding the interpretation of the Ly-$\alpha$ line-center data observed by CLASP
we will not aim at performing a point-by-point inversion of the atmospheric physical quantities. Instead, we will use the information contained in all the pixels along the CLASP spectrograph slit in order to determine the parametrization $\vec{\hat\theta}$ of suitable statistical models of the solar atmosphere that maximizes the marginal posterior of Eq.~(\ref{eq:finpost}). In other words, our first step will be to estimate some global properties of the quiet Sun atmosphere observed by CLASP (e.g., the average field strength and the degree of corrugation of the chromosphere-corona transition region). 

\subsection{Maximum Likelihood Estimation}

The general Eq.~(\ref{eq:finpost}) allows us to impose a reasonable prior distribution for the hyperparameters.
In this paper, for the hyperparameters of our CTR model we assume a flat prior distribution, $p(\vec\theta)\propto 1$, and we therefore restrict our analysis to the study of the likelihood $\mathcal{L}$.

Our goal is to find the estimator
\begin{equation}
\hat{\vec\theta}=  \underset  {\vec\theta\in\Theta}   {{\rm arg}\,{\rm max}\;}   \mathcal{L}(\vec\theta|\vec S^{\rm obs})\;,
\end{equation}
which maximizes the likelihood function, i.e., it corresponds to the model with the strongest observational evidence. Given two model parametrizations $\vec\theta_A$ and $\vec\theta_B$, the likelihood ratio
\begin{equation}
\Lambda(\vec\theta_A,\vec\theta_B|D) = \frac{\mathcal{L}(\vec\theta_A|\vec S^{\rm obs})}{\mathcal{L}(\vec\theta_B|\vec S^{\rm obs})}
\end{equation}
measures the strength of evidence of $\vec\theta_A$ with respect to $\vec\theta_B$.
In the following, the maximum of the likelihood function corresponding to a particular observation will be denoted by $\mathcal{L}_{\rm max}=\mathcal{L}(\hat{\vec\theta}|\vec S^{\rm obs})$ and the likelihood ratio
\begin{equation}
\Lambda=\frac{\mathcal{L}}{\mathcal{L}_{\rm max}}
\label{eq:likrat}
\end{equation}
will be used to quantify the model fitness.

\subsection{Maximum Likelihood Estimation Using the Corrugated Transition Region Model}
\label{ssec:mle}

\begin{figure}
\centering
\begin{tabular}{c}
\includegraphics[width=\columnwidth]{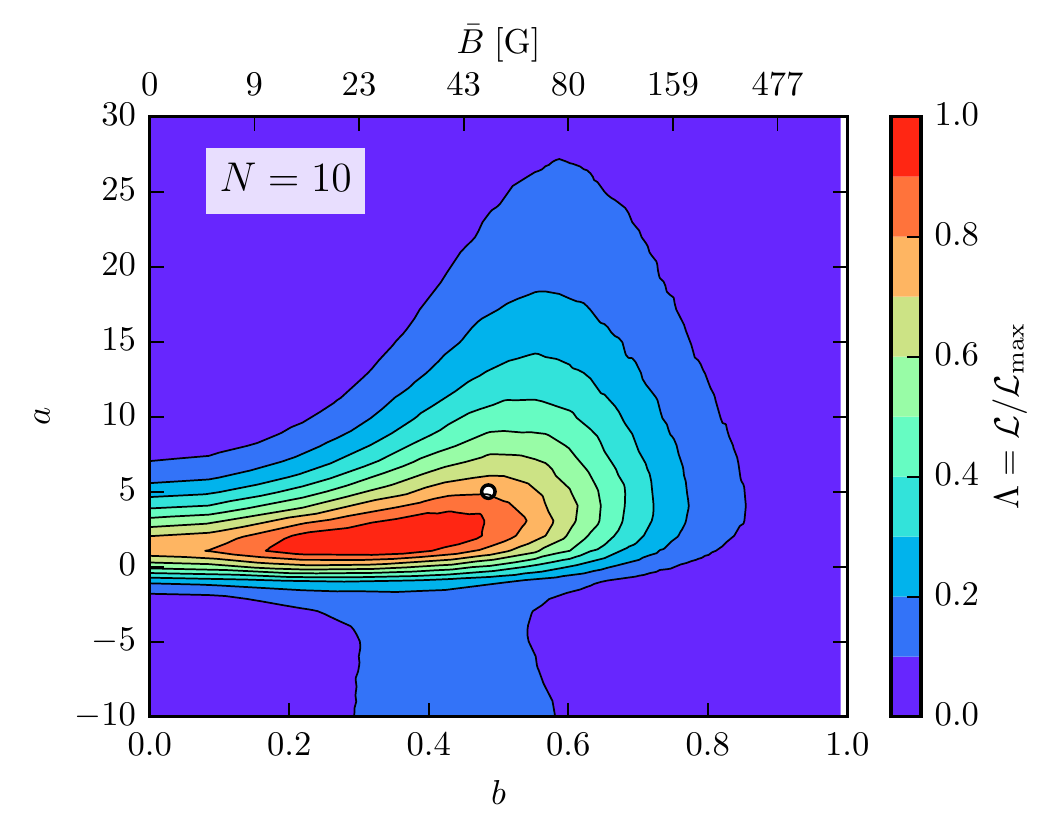} \\
\includegraphics[width=\columnwidth]{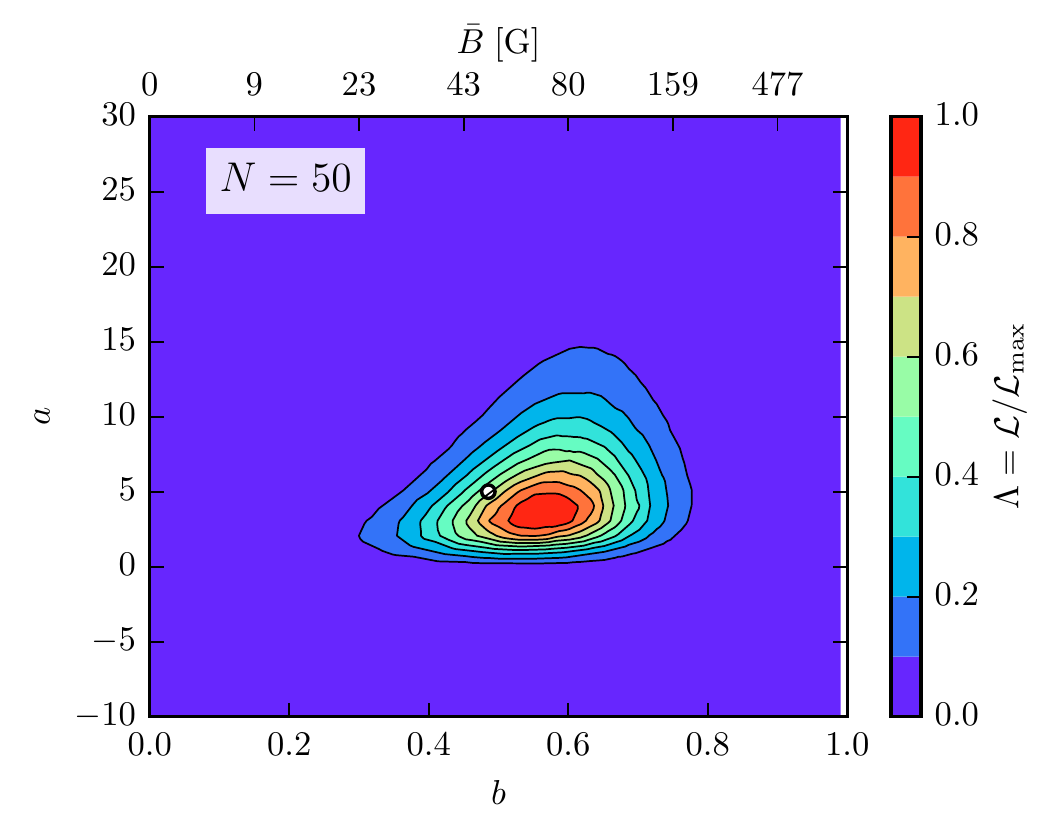} \\
\includegraphics[width=\columnwidth]{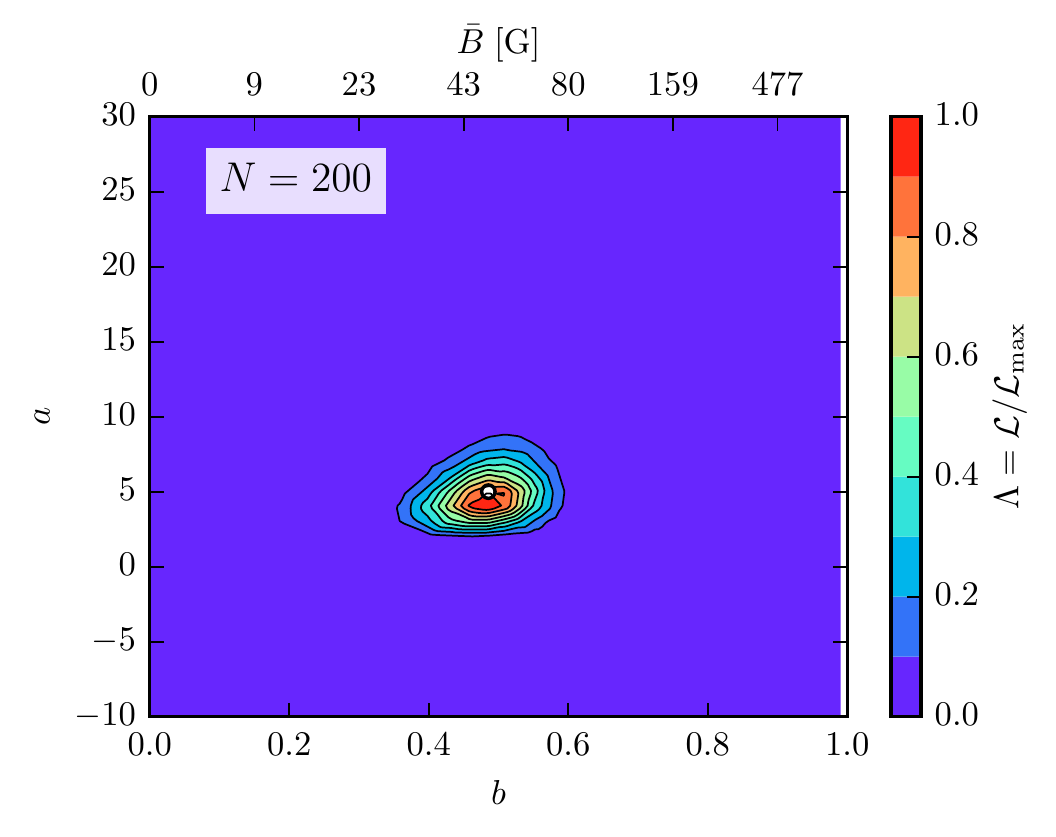}
\end{tabular}
\caption{
Maps of the likelihood ratios (see Eq.~\ref{eq:likrat}) corresponding to the three particular data realizations of Fig.~\ref{fig:psalc}. From top to bottom, the number of observed points is 10, 50, and 200. The white dot at $a=5$, $\bar B=50$\,G ($b\approx 0.5$) indicates the true model parameters. For $N=200$, the estimator, i.e., the point of the maximum likelihood, is already very close to the true value.
}
\label{fig:mle}
\end{figure}

In this section, we show a practical application of the Maximum Likelihood Estimation (MLE) method using synthetic data corresponding to the CTR model introduced in Appendix~\ref{app:ctr}. To this end, we assume that the true quiet solar chromosphere corresponds to our CTR model with the particular value of parameters, $\vec\theta_0=\{a_0,\bar B_0\}=\{5,50\,{\rm G}\}$. We use this model to generate random realizations of the Stokes parameters along the imaginary spectrograph slit oriented radially from near the solar limb towards the disk center. The above-mentioned data play the role of an actual observation and we will show how to infer the parameters of the model by applying the MLE.

Fig.~\ref{fig:psalc} shows three different realizations (i.e., synthetic observations) of a CLASP-like experiment with different numbers of data points affected by gaussian noise. It is important to realize that the accuracy of the MLE depends sensitively on the amount of available data. Under the assumption that the real solar atmosphere can be exactly described by the CTR model, then in the limit of an infinite number of data points, the MLE provides exact results. Finite datasets lead to approximate results. For this reason, it is important to study the accuracy of the method with respect to $N$.

Without loss of generality, we use the approximation from Section~\ref{ssec:distrps} for decoupling the Stokes parameters,
\begin{equation}
p^\mu(\vec S|\vec\theta) = p^{\mu}(q|\vec\theta)p^{\mu}(u|\vec\theta)\;.
\end{equation}
This factorization simplifies the practical calculations and our numerical tests show that it does not significantly affect the results at least in the case of the CTR model. The practical calculation of the likelihood of model hyperparameters $\vec\theta=\{a,\bar B\}$ requires to evaluate the integral on the r.h.s. of Eq.~(\ref{eq:llii}) for every pixel along the spectrograph slit. 
The PDFs of $q$ and $u$ can be approximated by normalized histograms of these quantities constructed from random samples of the TR realization using the CTR model. For any given parametrization, $\vec\theta$, the distribution of the normal vectors of the TR is given by Eq.~(\ref{eq:nvdist}) and the distribution of magnetic field strengths follows from Eq.~(\ref{eq:bdist}). 
The polarization of the emergent radiation can then be efficiently calculated using the methods of Appendix~\ref{app:ctr} and incorporated into the histogram of the polarization signals. 

Since the number of free parameters is only two, it is feasible to efficiently calculate a dense grid of models covering the relevant part of the parameter space and to determine the estimator $\hat{\vec\theta}$. If the number of free parameters is larger, more advanced methods such as the Markov chain Monte Carlo method can be considered.

As the magnetic field strength reaches the saturation limit of the Hanle effect ($\bar B\gg B_H$) the polarization signals are no longer sensitive to the value of $\bar B$. In order to be able to visualize the infinite range of magnetic field intensities, we use the parameter
\begin{equation}
b=1-\left(1+\frac{\bar B}{B_H}\right)^{-1}
\end{equation}
instead of $\bar B$.
The likelihood ratio corresponding to the slit realizations of Fig.~\ref{fig:psalc} are shown in Fig.~\ref{fig:mle}.
The parameter $b$ on the horizontal axis is equal to zero for $\bar B=0$\,G, it is $b=0.5$ for $\bar B=B_H$, and $b= 1$ for $\bar B=\infty$. It is clear that as the number of data points in the observation increases above about $10^2$, the maximum likelihood estimation becomes very close to the true value and the uncertainty of the inferred parameters significantly decreases.

\subsection{Note on the Limited Spatial Resolution of the Observations}
\label{ssec:resol}

The observed data are always affected by the finite spatial and spectral resolution of the instrument, and the limited resolution affects the statistical distibution of the observed signals. In the analysis of the data, one needs to take this fact into account and to apply the correct degradation procedure equivalent to that of the telescope and spectrograph to the synthetic spectra before constructing the $p^\mu(\vec S|\vec\theta)$ functions.

This cannot be done in the CTR model because this simple model does not provide any information about the local spatial variability of the Stokes parameters in the FOV (see \ref{ssec:ctrlim}). However, such information is available in 3D~MHD models in which one has to convolve the theoretical spectra with the point-spread function of the instrument and to degrade the spectra in the wavelength space. In our following paper, we apply the statistical inference method described in this paper to the CLASP observations.

\section{Concluding comments}
\label{sec:concl}

Recently, unprecedented spectropolarimetric observations of the hydrogen Lyman-$\alpha$ line of the solar disk radiation have been provided by the CLASP sounding rocket experiment. The $Q/I$ and $U/I$ line-center signals observed along the spatial direction of the (radially oriented) spectrograph slit encode key information on the magnetic field and geometrical complexity of the corrugated layer that delineates the chromosphere-corona transition region. With only one spectral line it is not possible to determine the magnetic field of the solar transition region, unless the geometrical complexity of its defining corrugated layer is known beforehand. However, it should be possible to constrain the mean field strength and geometrical complexity via the application of a suitable statistical inference method. 

The aim of this paper has been to propose a statistical inference method, based on the concept of maximum likelihood, which we consider suitable for interpreting the CLASP observations. Here we have illustrated this method by applying it to the theoretical $Q/I$ and $U/I$ line-center signals resulting from a simple line formation model (the corrugated transition region model) that we have introduced for qualitatively understanding the CLASP observations. In our next paper (Trujillo Bueno et al. 2018; in preparation) we will apply the statistical inference method developed here in order to interpret the CLASP data themselves and estimate global properties of the quiet Sun atmosphere  
by means of more realistic models of the chromosphere-corona transition region resulting from 3D numerical simulations.


\acknowledgements

The CLASP team is an international partnership between NASA Marshall Space Flight Center, 
National Astronomical Observatory of Japan (NAOJ), Japan Aerospace Exploration Agency 
(JAXA), Instituto de Astrof\'{i}sica de Canarias (IAC) and Institut d'Astrophysique Spatiale; 
additional partners are the Astronomical Institute ASCR, Istituto Ricerche Solari Locarno (IRSOL), 
Lockheed Martin and University of Oslo. The US participation was funded by NASA Low Cost Access to Space (Award 
Number 12-SHP 12/2-0283). The Japanese participation was funded by the basic research 
program of ISAS/JAXA, internal research funding of NAOJ, and JSPS KAKENHI Grant 
Numbers 23340052, 24740134, 24340040, and 25220703. The Spanish participation was 
funded by the Ministry of Economy and Competitiveness through project AYA2010-18029 
(Solar Magnetism and Astrophysical Spectropolarimetry). The French hardware participation 
was funded by Centre National d'Etudes Spatiales (CNES). Moreover, we acknowledge the supercomputing grants provided by the Barcelona Supercomputing Center (National Supercomputing Center, Barcelona, Spain), as well as the financial support received through grant \mbox{16--16861S} of the Grant Agency of the Czech Republic, project \mbox{RVO:67985815} of the Czech Academy of Sciences, and projects \mbox{AYA2014-60476-P} and \mbox{AYA2014-55078-P} of the Spanish Ministry of Economy and Competitiveness. JS is grateful to the Visiting Researchers Programme of the Jes\'us Serra Foundation for financing a four months working visit at the IAC. Finally, JTB acknowledges the funding received from the European Research Council (ERC) under the European Union's Horizon 2020 research and innovation programme (ERC Advanced Grant agreement No 742265).

\appendix
\section{Corrugated transition region model (CTR)}
\label{app:ctr}

\subsection{Definition}

\begin{figure}
\centering
\includegraphics[width=7cm]{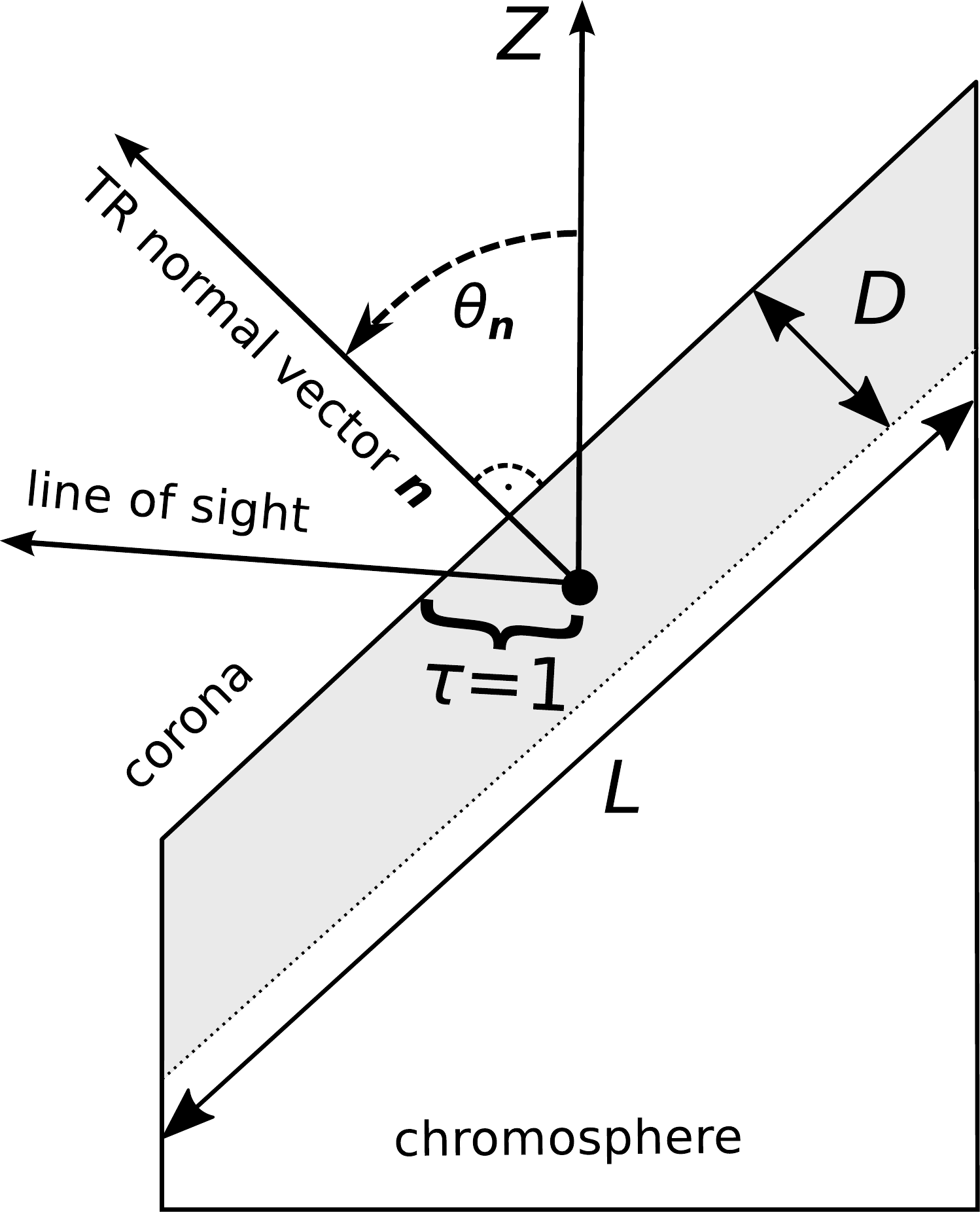}
\caption{
Motivation for the CTR model. The physical quantities of the transition region of geometrical thickness $D$ in which the Ly-$\alpha$ line core forms vary much faster along the local TR normal vector $\vec n$ than along the perpendicular direction. Therefore, we approximate the line-core formation region using an inclined plane-parallel model whose normal vector $\vec n$ makes an angle $\theta_n$ with the respect to the local vertical direction $Z$. The quantity $L$ is of the order of the local radius of curvature of the TR.
}
\label{fig:cartoon}
\end{figure}

The corrugated transition region (CTR) model is a simple approximation to the full 3D problem of formation of the Ly-$\alpha$ line core. Inspired by the peculiarities of the Ly-$\alpha$ formation that have been discussed in Sect.~\ref{sec:3d}, the CTR model assumes that the TR can be {\em locally} approximated by a thin slab whose normal vector is inclined with respect to the local vertical (see Fig.~\ref{fig:cartoon}). For our purposes, we assume that the thermal structure of the slab is that of the semi-empirical FAL-C model \citep{fontenla93}. Since in the CTR model the problem becomes effectively 1D instead of 3D, it allows us to greatly speed up the calculations while retaining some of the crucial properties of the symmetry breaking mechanism of the full 3D problem.

In the CTR model, the physical state of the TR at any given point and at any instant of time can be described by set of parameters $\vec\xi\equiv\left\{\mu_B, \chi_B, B, \mu_n, \chi_n\right\}$, where $\mu_B=\cos\theta_B$ and $\chi_B$ are the cosine of the inclination and the azimuth of the magnetic field vector, respectively, $B$ is the magnetic field strength, and $\mu_n=\cos\theta_n$ and $\chi_n$ are the cosine of the inclination and azimuth of the normal vector of the TR, respectively. For a given set of parameters $\vec\xi$, it is easy to synthesize the emergent Stokes profiles for any LOS using  a 1D non-LTE radiative transfer code and applying the necessary rotations of the reference frame (see below).

The statistical model of the quiet Sun atmosphere introduced in Sect.~\ref{ssec:distrps}, allows us to randomly sample the local atmospheric parameters $\vec\xi$ from a probability density distribution $p(\vec\xi|\vec\theta)$ given the model parameters $\vec\theta$ that play the role of hyperparameters for the local physical quantities. In the CTR model, we define the joint PDF in a factorized form 
\begin{equation}
p(\vec\xi|\vec\theta)=p(\mu_n|\vec\theta)p(\chi_n|\vec\theta)p(\mu_B|\vec\theta)p(\chi_B|\vec\theta)p(B|\vec\theta)\;.\label{eq:factor}
\end{equation}
It is evident that $p(\chi_n|\vec\theta)$ and $p(\chi_B|\vec\theta)$ must be uniform in the interval $[0,2\pi)$. In order to keep the model simple, we futhermore assume that the magnetic field vector is distributed isotropically, hence $p(\mu_B|\vec\theta)$ is uniform in $[-1,1]$. Therefore, the resulting distribution reads
\begin{equation}
p(\vec\xi|\vec\theta)=\frac{1}{8\pi^2} p(\mu_{\vec n}|\vec\theta)p(B|\vec\theta)\;.
\end{equation}

\begin{figure}
\centering
\includegraphics[width=9cm]{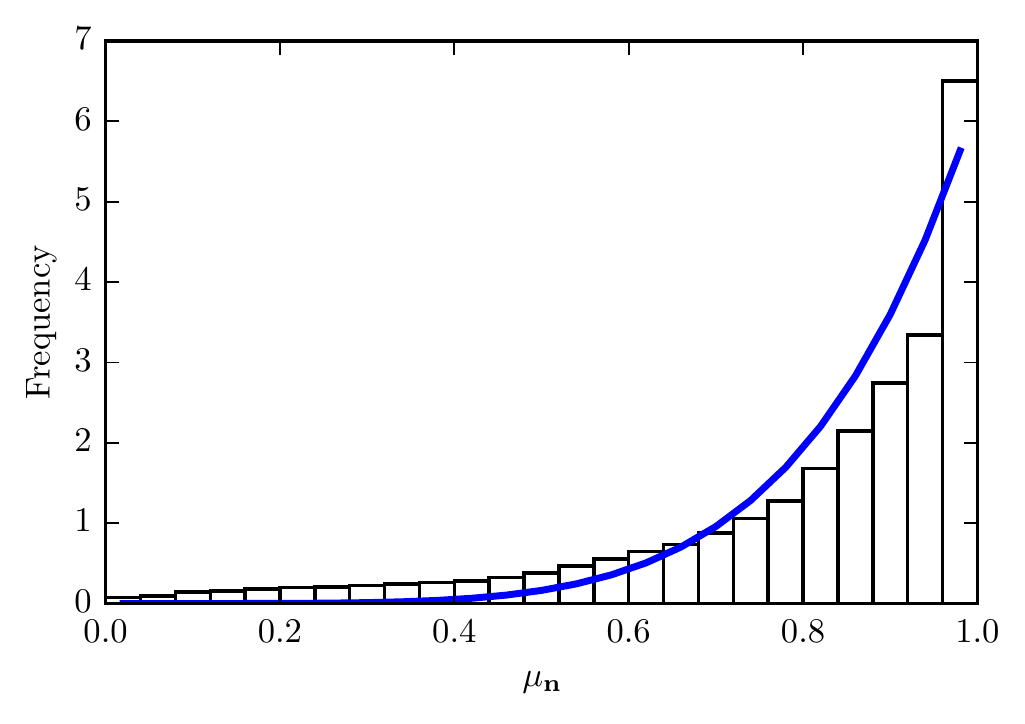}
\caption{
Histogram of the cosines of the normal vectors, $\mu_{\vec n}$, in the 3D MHD model of \cite{carlsson16}  
and its best least-squares fit with a distribution of Eq.~(\ref{eq:nvdist}) with $a\approx 5$.
}
\label{fig:normal3d}
\end{figure}

In this paper, we assume the PDF of $\mu_n$ to have the form
\begin{equation}
p(\mu_n|a) = \left\{
  \begin{array}{rl}
    (1+a)\mu_n^a & \qquad{\rm for}\quad a \ge 0 \\
    (1-a)(1-\mu_n)^{-a} & \qquad{\rm for}\quad a<0
  \end{array}
\right.,
\label{eq:nvdist}
\end{equation}
and we assume $\mu_n\in[0,1]$, i.e., we do not allow the TR to be oriented towards the solar interior.
The hyperparameter $a\in(-\infty,\infty)$, hereafter the {\em corrugation parameter}, determines whether the TR normal vector is predominantly vertical ($a>0$) or predominantly horizontal ($a<0$). The limiting cases of interest are the well-known case of a plane-parallel atmosphere ($a=\infty$) and the case of an atmosphere composed of vertically oriented plasma structures ($a=-\infty$). In the  $a\approx 0$ case, the distribution of $\mu_{\vec n}$ is that of highly chaotic orientations of the TR. As an illustration of the adequacy of the distribution \ref{eq:nvdist}, we show in Fig.~\ref{fig:normal3d} the histogram of $\mu_n$ obtained from a 3D model snapshot of an enhanced network region resulting from a MHD simulation by \cite{carlsson16}. The distribution qualitatively agrees with the functional form of Eqs.~(\ref{eq:nvdist}) with the best fit obtained for $a\approx 5$ (see the solid curve).

The CTR distribution of magnetic field intensity, $B$, is given by the Rayleigh distribution
\begin{equation}
p(B|\bar B)=\frac{\pi B}{2\bar B}\exp\left\{-\frac{\pi B^2}{4\bar B^2}\right\}\;,
\label{eq:bdist}
\end{equation}
parametrized by the average magnetic field value, $\bar B$, which is the second hyperparameter of the CTR model.\footnote{In more realistic atmospheric models, one should instead assume a log-normal distribution. However, the Rayleigh distribution depends on just one free parameter, which is advantageous in the inference problem.}

We note that the factorization in Eq.~(\ref{eq:factor}) is possible only if the involved quantities are independent. In reality, it is very likely that the orientation of the magnetic field vector is closely related to the orientation of the observed plasma structures in the transition region \citep[see, e.g.,][]{vourlidas10}. In our academic CTR model, for simplicity we do not consider such correlations. In the next paper of this series we will show that our inference of information from the CLASP line-center data is based on more realistic statistical model of the solar TR, based on parametrized 3D models having different levels of magnetization and geometrical complexity.

\subsection{Calculation of the Emergent Polarization Signals}

The hyperparameters $\vec\theta=\{a,\bar B\}$ define the model for the random atmosphere realizations. Since we use the FAL-C semi-empirical model for our calculations, the emergent Stokes parameters need to be calculated numerically applying a non-LTE radiative transfer code. It would be very time consuming to perform a non-LTE calculation for every possible combination of the normal and magnetic field vectors. Therefore, we use the Eddington-Barbier approximation to speed up the calculations. We consider an unmagnetized 1D plane-parallel atmosphere. For each LOS defined by its $\mu$ value, the emergent $Q/I$ signal at the center of Ly-$\alpha$ can be approximated by \citep{jtb-stepan-casini11}
\begin{equation}
\frac{Q(\mu)}{I(\mu)}\approx \frac{1}{2\sqrt 2}(1-\mu^2) \frac{J^2_0(\mu)}{J^0_0(\mu)}\;,
\label{eq:eb}
\end{equation}
where $J^K_Q$ are the multipolar components of the radiation field tensor at the height along the LOS where the line-center optical depth $\tau$ is unity. For each $\mu$ value, the self-consistent non-LTE calculation provides the line-center $Q/I$. From Eq.~(\ref{eq:eb}), we calculate the corresponding $J^2_0$ and $J^0_0$ values.\footnote{We point out that the use of these values for $J^2_0/J^0_0$ provide better accuracy than the use of those at $\tau=1$.}

In order to calculate the emergent signals from the inclined TR at the point being considered, we use the formalism of the irreducible spherical tensors.
Given the components of a general spherical tensor $[T^K_P]_{\rm old}$ in an `old' reference frame, its components in a `new' frame are given by the transformation
\begin{equation}
[T^K_Q]_{\rm new}=\sum_{P} [T^K_P]_{\rm old} \mathcal{D}^K_{PQ}(R)\;,
\label{eq:jrot}
\end{equation}
where $R$ is the rotation that brings the old frame to the new one and $\mathcal{D}^K_{PQ}(R)$ is the corresponding rotation matrix \citep[see Sect.~2.7 of][]{LL04}.

Calculation of the emergent fractional polarization from the rotated TR involves several transformations of the coordinate systems. In the following, we use three different reference frames: (1) the laboratory frame, $L$, whose positive $Z$ axis is parallel to the local vertical in the atmosphere; (2) the TR frame, $T$, whose positive $Z$ axis is parallel to the normal vector of the TR; (3) the magnetic field reference frame, $M$, whose positive $Z$ axis is parallel to the local magnetic field vector. The algorithm to calculate the emergent fractional polarization for a particular LOS, orientation of the TR normal vector, and arbitrary magnetic field, is as follows:
\begin{enumerate}
\item
Calculate the cosine of the inclination of the LOS vector, $\vec{\hat l}$, in the TR frame: $[\mu]_T=\vec{\hat l}\cdot\vec n$.
\item
For $[\mu]_T$, we obtain the value of $[Q/I]_T$ from the CLV curve of FAL-C (this curve is calculated by solving numerically the 
full non-LTE radiative transfer problem in the absence of magnetic fields).
\item
From Eq.~(\ref{eq:eb}), we get $[J^2_0/J^0_0]_T$.
\item
We express the $[J^K_Q]_M$ tensor in the magnetic field reference frame (hereafter, the $M$-frame) 
by rotating the reference frame using Eq.~(\ref{eq:jrot}).
\item
We solve the statistical equilibrium equations in the magnetic field reference frame to obtain $[\rho^K_Q]_M$ (i.e., the multipolar components of the atomic density matrix of the only level of Ly-$\alpha$ that contributes to its scattering polarization, namely its upper level 
with $J=3/2$). This is particularly easy in the $M$-frame by using Eq.~(10.27) of \citet{LL04}.
\item
Express the atomic density matrix in the laboratory frame $[\rho^K_Q]_L$ by rotating the reference frame from $M$ to $L$.
\item
Calculate the emergent Stokes parameters in the laboratory frame.
\end{enumerate}

\subsection{CTR Model Limitations}
\label{ssec:ctrlim}

The CTR model  helps us to easily understand some of the key aspects of the Ly-$\alpha$ line core polarization in the solar TR. Thanks to its simplicity, it is also useful for demonstrating the general statistical inference method that will be applied in our next paper. Below, we summarize the main deficiencies of the CTR model which makes it of limited use for practical applications.

The CTR model is based on only one atmospheric component, FAL-C. Therefore it cannot take into account differences between network and internetwork and, in general, it cannot be used to fit the line-center  intensity. This problem can be partially solved by using 1D multi-component models, but since the CTR model cannot be used in practice for number of other reasons, such generalization would probably be inappropriate.

With the CTR model, we cannot account for the limited spatial resolution of the observations. 3D radiative transfer calculations show that the fractional polarization amplitudes can reach up to several percent in the case of observations with very high spatial resolution \citep{stepanjtb15}. The spatial smearing of the signals leads to a significant decrease of the amplitudes. Accidentally, the amplitudes of the $Q/I$ and $U/I$ CLASP signals are comparable to those corresponding to the CTR model, but the CLASP amplitudes are already significantly affected by the limited spatial resolution and the true solar signals have, therefore, necessarily higher amplitudes.

While the CTR model is an intuitive approximation to reality, the symmetry breaking at a given point in the TR is always affected by the illumination from distant regions that reaches that point through the optically thin gaps of coronal plasma \citep[see Fig.~3 of][]{stepanjtb15}. In addition, the local curvature of the TR is often not negligible and it produces additional symmetry breaking.



\end{document}